\documentclass{aa}
\usepackage{graphicx}
\usepackage{epsfig}
\usepackage{bm}

\usepackage{multirow}
\usepackage{orcidlink}
\usepackage[varg]{txfonts}
\usepackage{url}
\usepackage[normalem]{ulem}
\usepackage{color}
\usepackage{hyperref}
\usepackage{color}
\usepackage[switch]{linenoaa} 
\linenumbers
\usepackage{natbib}
\defcitealias{Bobrikova25}{B25}

\newcommand{\diff}{{\rm d}}
\newcommand{\ixpe}{\textit{IXPE}\xspace}
\newcommand{\istar}{i_*}
\newcommand{\phinew}{\phi}
\newcommand{\thetanew}{\theta}

\newcommand{\iorb}{i}

\newcommand{\xnew}{\unit{x}_*}
\newcommand{\ynew}{\unit{y}_*}
\newcommand{\znew}{\unit{z}_*}

\newcommand{\unit}[1]{\mbox{\boldmath $\hat{#1}$}}

\begin{document}

\title{Breaking the symmetry: X-ray polarization from a spreading layer on a tilted weakly magnetized neutron star} 

\titlerunning{Polarization from a tilted spreading layer}
\authorrunning{Bobrikova, A., et al.}

\author{Anna~Bobrikova\inst{1}\corrauth{anna.a.bobrikova@utu.fi}\orcidlink{0009-0009-3183-9742}
\and Alexandra Veledina\inst{1,2}\orcidlink{0000-0002-5767-7253} 
\and Juri~Poutanen\inst{1}\orcidlink{0000-0002-0983-0049}
\and Vladislav~Loktev\inst{3}\orcidlink{0000-0001-6894-871X}
}

\institute{Department of Physics and Astronomy, FI-20014 University of Turku, Finland 
\and Nordita, Stockholm University and KTH Royal Institute of Technology, Hannes Alfv\'ens v\"ag 12, SE-10691 Stockholm, Sweden
\and School of Mathematics, Statistics, and Physics, Newcastle University, Newcastle upon Tyne NE1 7RU, UK 
}

\date{\today}

\abstract{When modeling the emission from an X-ray binary system, we often assume that the rotational axis of the compact object is aligned with that of the binary system. While often considered irrelevant for spectroscopic observations, the orientation of the binary system is crucial for polarimetric properties. With the recent advent of X-ray polarimetric observations enabled by the \textit{Imaging X-ray Polarimetry Explorer} (\ixpe), indications of a possible misalignment of the neutron star rotational axis with respect to the binary axis have been reported. Following these indications, we present an analytical model for the polarimetric properties of emission arising from a spreading layer on an inclined weakly magnetized neutron star (WMNS). We discuss the expected polarimetric properties and study the impact of geometrical parameters of the system on the polarization of WMNS emission. We compare our results with those for aligned systems. Finally, we discuss applications of the model to recent and upcoming X-ray polarimetric observations with \ixpe and mention the suggestions for further development of the model.
}

\keywords{methods: analytical -- polarization -- stars: neutron -- X-rays: binaries}

  \maketitle
\nolinenumbers


\section{Introduction}

In an X-ray binary with a weakly magnetized neutron star (WMNS), the accretion flow is barely affected by the magnetosphere of the neutron star (NS) and is instead shaped by gravity. Once the companion star fills its Roche lobe, matter from the outer layers of the companion star is transferred towards the NS. The matter transferred through the inner Lagrangian point carries a high specific angular momentum; it circulates around the NS, forming an accretion disk \citep{ShakuraSunyaev73}. In WMNS systems, the mass accretion rate is so high that the accretion disk can push the magnetosphere towards the NS surface or even burying it beneath the NS surface. Consequently, the accretion disk is either mildly affected by the magnetosphere or reaches the NS surface unaffected by the magnetic field; in either scenario, matter accretes onto the equatorial region of the NS rather than onto the magnetic field poles as in pulsars. As the accretion disk reaches the NS surface, the matter is forced to decelerate rapidly from near-Keplerian velocities in the accretion disk to the rotational velocity of the NS surface before settling onto the NS. A so-called boundary layer \citep[BL; see, e.g.,][]{Shakura88} is formed. The accreted matter can subsequently spread over the NS surface, forming a spreading layer \citep[SL, ][]{IS99}.\footnote{We note here that the terms BL and SL have been conflated in the literature; in this paper, we use BL to denote the boundary region between the accretion disk and the NS surface, and SL for the extension of the accretion flow onto the NS surface.}

The spectrum of WMNS systems is traditionally decomposed into two main components: the softer disk emission and the harder emission from the BL/SL. 
Additional spectral features include reflection of the harder component from the disk and scattering of the emission in the wind above the disk. 
With spectroscopy alone, the disk and BL/SL components are often hard to distinguish. 
Timing analysis helps distinguish these components, since BL/SL emission is expected to vary on much faster timescales. 
For instance, quasi-periodic oscillations (QPOs) have been shown to be associated with the BL/SL region \citep{GRM03, RG06}. 
However, timing cannot determine the exact geometry of the accretion flow or distinguish between the BL and SL configurations. 

Polarimetry holds promise to investigate the geometry of binary systems. With the launch of the \textit{Imaging X-ray Polarimetry Explorer} \citep[\ixpe;][]{Weisskopf2022} in December 2021, it became possible, for the first time, to measure with high precision the polarization degree (PD) and polarization angle (PA) of X-ray emission. 
In addition to providing information on the geometrical configuration of the system, polarimetric properties carry information about the processes affecting the emission as it travels from the source to the observer, since X-ray emission becomes polarized through scattering, reflection, or interaction with the magnetic field. 
In recent years, more than twenty WMNS systems have been observed with \ixpe, and observations of many of these systems have provided constraints on the source geometry. 

Since the early stages of X-ray astronomy, various theoretical models for the expected polarized emission of binary systems have been developed in preparation for X-ray polarimetric observations. For instance, the polarimetric properties of accretion disks were studied in \cite{Rees75} based on pioneering work of \cite{Cha60} and \cite{Sob63}, under the assumption of electron-scattering-dominated disk atmospheres. 
Relativistic effects were later incorporated into the model by, e.g., \cite{Connors80}, \cite{2008MNRAS.391...32D}, and \cite{LVP22}. 
Currently, for linear polarization in the \ixpe energy range, the expected PD is 2\% for moderate inclination, rising to 6\% for $i=80\degr$ \citep{LVP22}. Polarization is expected to be aligned with the accretion disk plane. 

For the BL/SL, polarimetric expectations strongly depend on the geometry of the region. 
Both pioneering analytical works \citep{Sunyaev1985, LS1985} and modern numerical simulations, e.g., \cite{2022MNRAS.514.2561G}, predict that emission arising from a BL configuration of the accretion flow is polarized in the plane of the accretion disk, whereas for the SL configuration, the emission is expected to be polarized perpendicular to the disk plane. 
In recent analytical modeling, \citet[][hereafter B25]{Bobrikova25} calculated that the PD of the SL is not expected to exceed 1.5\% and that the emission is polarized in the plane perpendicular to the accretion disk, with minor variations depending on the energy and geometrical configuration of the system. 
Similar conclusions are presented in \cite{2024A&A...684A..62F}. 

Both scattering in the wind \citep{Nitindala2025} and reflection from the accretion disk \citep{LS1985,Matt1993, Poutanen96} are expected to boost the total polarization of the WMNS emission. However, the contributions of reflection component to the total flux are usually small, and the scattered fraction is low for the systems observed from low and medium inclinations \cite{Nitindala2025}, so the general expected PD of WMNS systems remains relatively low. 

However, recent \ixpe observations have shown that for many sources the observed PD values significantly exceed theoretical expectations \citep[see, e.g.,][for reviews]{2024Galax..12...43U, 2025AN....34640126D,2025A&A...699A.230G}. In several cases, at higher energies, where the Comptonized component is expected to dominate the flux, the PD exceeded 10\%. 
Among possible explanations for the high PD, scattering in the wind and reflection from the disk are usually mentioned \citep[see, among many others,][]{2025A&A...699A.230G}; however, not all the systems are known to have strong winds or to show signs of reflection. For instance, \mbox{GX~5$-$1} is known for a lack of reflection features \citep{Homan2018} and does not show signs of a strong wind in X-ray spectroscopy, yet the observed PD exceeds 4\% at higher energies \citep{Fabiani24}. One possible explanation for the unexpectedly high PD in these systems is a breaking of the axial symmetry. 

Another known issue is the PA of the continuum components: although theoretically the PA of the Comptonized component is expected to be aligned with or orthogonal to that of the disk, for several sources, a PA difference of about $60\degr$ has been observed. Notably, the young X-ray binary \mbox{Cir~X-1} was one such source, and in this source, a misalignment between the rotational axis of the NS and the binary axis was suggested as a possible explanation \citep{Rankin2024}, since hints of misalignment such as precession of the accretion disk and twisted jets had already been observed in this system \citep{2019MNRAS.484.1672C, 2025MNRAS.544L..37C}. 

Here, we introduce an analytical model of polarized SL emission from a misaligned NS. In Sect.~\ref{sec:theory} we present the model, describing the physical properties of the SL emission, its propagation, and the geometrical configuration of the system. In Sect.~\ref{sec:Results} we present the results of our study of the impact of various system parameters on the observed emission. We discuss the application of our model to previous \ixpe observations in Sect.~\ref{sec:IXPE}. Finally, we conclude in Sect.~\ref{sec:Summary}. 

\section{Model setup} 
\label{sec:theory} 

\begin{figure}
\begin{center}
\includegraphics[width=0.95\linewidth]{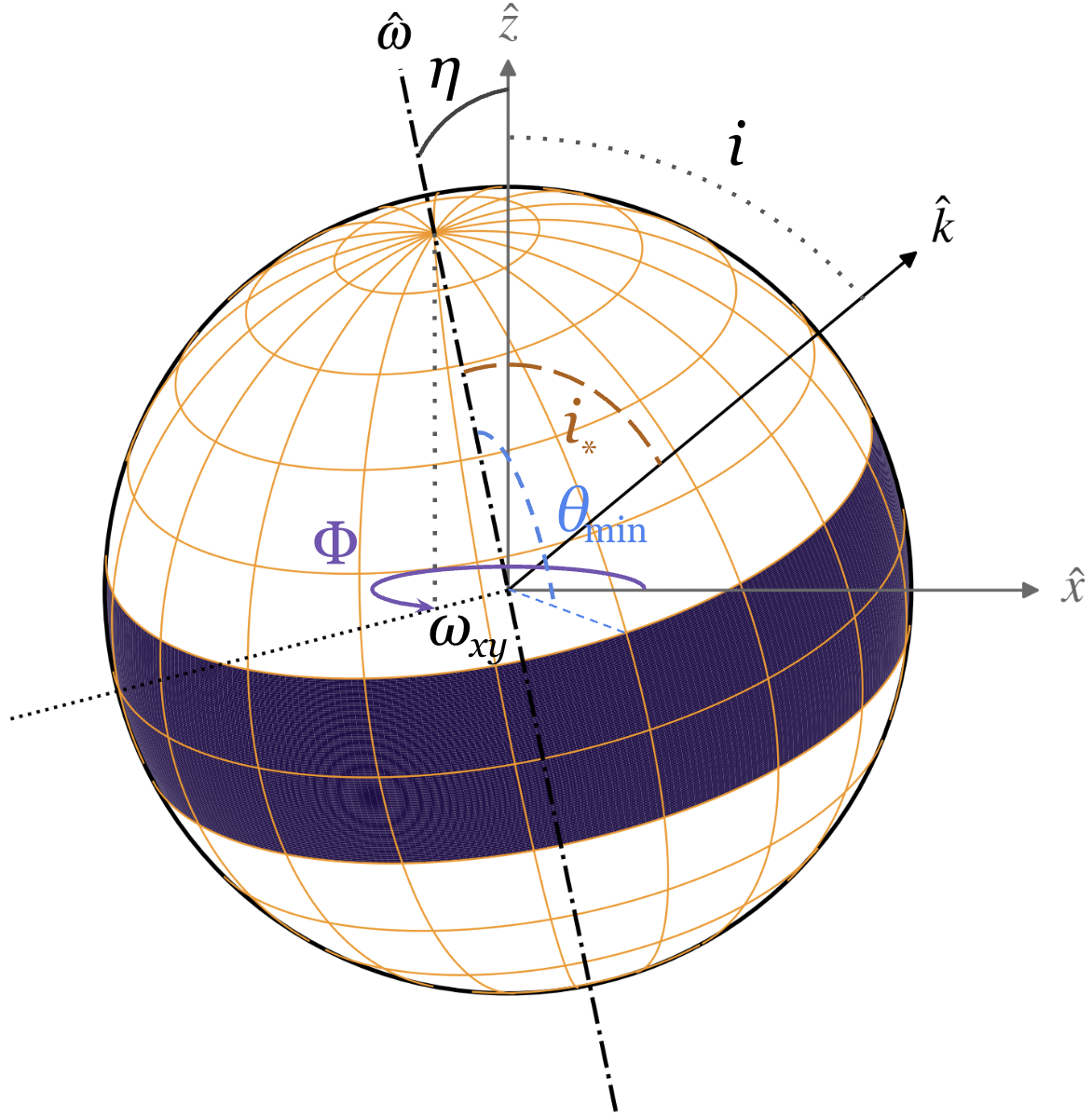}
\caption{Geometry of the problem. The rotational axis of the NS $\omega$ is misaligned relative to the rotational axis of the binary system ($z$) by an angle $\eta$. 
The rotational axis of the NS forms the azimuthal angle $\Phi$ (marked with a purple solid line) with the direction to the observer ($\unit{k}$).
The inclinations of the binary system, $\iorb$, and of the NS, $\istar$, are indicated by black dotted and brown dashed lines, respectively. 
The SL is shown in dark purple and is confined between co-latitudes $\theta_{\rm min}$ (marked with a blue dashed line) and $\theta_{\rm max}=180\degr-\theta_{\rm min}$. The accretion disk lies in the $xy$ plane; it is omitted from the sketch for clarity.}
\label{fig:geom}
\end{center}
\end{figure}

The geometry of the system is shown in Fig.~\ref{fig:geom}. We consider a spherical NS of radius $R$ and mass $M$ with a SL centered at the NS equator and extending between the co-latitudes of the SL edges, $\theta_{\rm min}$ and $\theta_{\rm max}=180\degr-\theta_{\rm min}$. The matter in the SL flows counterclockwise with the equatorial velocity $\beta_{\rm eq}$; the velocity decreases towards the poles as described in Eq.~\eqref{eq:velocity}. The SL emission is assumed to be a (modified) blackbody at the constant effective temperature $T_{\rm eff}$ and is modeled as a semi-infinite electron-scattering-dominated plane-parallel slab. 

We assume that the rotation axis of the binary system is aligned with the $z$-coordinate axis, and the accretion disk lies in the $xy$-coordinate plane. The coordinates are chosen so that the unit vector in the direction of the observer $\unit{k}$ lies in the $xz$ plane and makes an angle $\iorb$ with the binary rotation axis, giving 
\begin{equation}
\unit{k} = (\sin \iorb, 0, \cos \iorb). 
\end{equation} 

We assume that the NS rotational axis $\unit{\omega}$ is misaligned with respect to the rotational axis of the binary system by an angle $\eta$. 
In the binary plane, the NS rotation axis forms an azimuthal angle $\Phi$ with the direction to the observer $\unit{k}$. 
In Fig.~\ref{fig:geom}, $\eta$ is measured from the positive $z$-axis, while $\Phi$ is measured in the $xy$-plane from the positive $x$-axis. Hence, the NS rotational axis is defined as follows:
\begin{equation} 
\unit{\omega}=(\sin \eta \cos \Phi, \sin \eta \sin \Phi, \cos \eta).
\end{equation} 
The NS inclination $\istar$, defined as the angle between the direction to the observer and the NS rotational axis $\unit{\omega}$, is given by 
\begin{equation}\label{eq:i-star}
\cos \istar = \unit{k} \cdot \unit{\omega}= \cos \iorb \ \cos\eta + \sin \iorb \sin\eta\cos\Phi.
\end{equation} 

To relate the current model to the axisymmetric case presented in \citetalias{Bobrikova25}, where the NS rotation axis was aligned with the orbital axis, we move to the coordinate system associated with the NS.
We then define all the necessary vectors in that new coordinate system, which we refer to as the ``NS frame'', and denote their coordinates with the subscript *. 
Here, $z_*$ is aligned with the NS rotational axis $\unit{\omega}= (0, 0, 1)_*$ and we define the coordinate axes as follows:
\begin{eqnarray}
\xnew &=& \ynew\times\unit{\omega}
= (\unit{k}-\unit{\omega}\cos \istar)/ \sin \istar, \nonumber \\
\ynew &=& \unit{\omega}\times\unit{k} / \sin \istar, \\
\znew &=& \unit{\omega}. \nonumber 
\end{eqnarray}
The direction to the observer in this system is 
\begin{equation}
\unit{k} =(\sin \istar, 0, \cos \istar)_* . 
\end{equation} 
Introducing the co-latitude $\thetanew$ and the azimuthal angle $\phinew$, which specify a point on the NS surface in that coordinate system, we can define the normal to the surface element:
\begin{equation}
\unit{n} = (\sin\thetanew \cos\phinew, \sin\thetanew \sin\phinew, \cos\thetanew)_*.
\end{equation} 
This geometry is used to calculate the polarized emission from the SL on an aligned NS using the formalism presented in \citet{Suleimanov20} and \citetalias{Bobrikova25}. For completeness, we provide the necessary formulae in Appendix~\ref{app:a}. 
We assume that the spectrum of the local emission can be described by a diluted blackbody 
\begin{equation}\label{eq:dilutbb}
I'_{E'} (\alpha')=a_{\rm es}(\alpha') \frac{1}{f^4_{\rm c}} B_{E'} (f_{\rm c}T_{\rm eff}) , 
\end{equation} 
where $B_{E'}$ is the Planck function, $f_{\rm c}$ is the color correction factor, and $a_{\rm es}(\alpha')=0.421 + 0.868 \cos \alpha'$ is an approximation to the angular distribution of radiation intensity from a semi-infinite, electron-scattering-dominated atmosphere \citep{1947ApJ...105..435C} as a function of the zenith angle $\alpha'$ in the local comoving frame of the gas.  
The polarized flux (Stokes vector) observed from an SL surface element $\diff S$ is given by
\begin{equation} 
\diff \bm{F}(E)= \left(\begin{array}{c} \diff F_I(E) \\ \diff F_Q(E) \\ \diff F_U (E) \end{array}\right) = \diff F_E \left(\begin{array}{c} 1 \\ P(\alpha')\cos2\chi \\ P(\alpha') \sin2\chi \end{array}
\right) ,
\end{equation} 
where $P$ is the PD at the zenith angle $\alpha'$, calculated using the approximation \citep{Cha60} for a semi-infinite electron-scattering-dominated atmosphere.
The PA $\chi$ is calculated using the formula derived in \citet{Poutanen2020} that accounts for corrections caused by the relativistic motion of the SL matter and light bending in the Schwarzschild metric. 

The misalignment must be included in the PA calculation. 
The PA is measured counterclockwise with respect to the orbital axis as $\chi_{\rm tot} = \chi+\chi_{\rm m}$, where the angle $\chi_{\rm m}$ is the position angle of the NS rotational axis measured eastward (i.e. counterclockwise) from the projection of the orbital axis onto the sky. 
To compute $\chi_{\rm m}$, we introduce the polarization basis related to the orbital coordinate system:
\begin{eqnarray}
\unit{e}_x^z&=&\frac{\unit{z}-\unit{k}\cos{\iorb}}{\sin{\iorb}}
=(-\cos{\iorb}, 0, \sin{\iorb}), \\
\unit{e}_y^z&=&\frac{\unit{k}\times\unit{z}}{\sin{\iorb}}=(0, -1, 0).
\end{eqnarray} 
The polarization basis related to the coordinate system of the misaligned NS is similarly defined as 
\begin{eqnarray}
\unit{e}_x^{*}&=&\frac{\unit{\omega}-\unit{k}\cos\istar}{\sin\istar} = \frac{1}{\sin\istar} \left(\begin{array}{c} \sin\eta\cos\Phi-\sin{\iorb}\cos\istar \\ \sin\eta\sin\Phi \\ \cos\eta - \cos{\iorb}\cos\istar \end{array}\right) , \\
\unit{e}_y^{*}&=&\frac{\unit{k}\times\unit{\omega}}{\sin\istar}. 
\end{eqnarray} 
Then, we calculate $\chi_{\rm m}$ using
\begin{eqnarray}
\cos{\chi_{\rm m}} &= &\unit{e}_x^{*}\cdot\unit{e}_x^z = \frac{\cos\eta\sin{\iorb} - \sin\eta\cos{\iorb}\cos\Phi}{\sin\istar}, \\
\sin{\chi_{\rm m}} &= &\unit{e}_x^{*}\cdot\unit{e}_y^z = -\frac{\sin\eta\sin\Phi}{\sin\istar}, \nonumber
\end{eqnarray} 
or
\begin{equation}
\tan \chi_{\rm m} = \frac{\sin\eta \sin\Phi}{\sin\eta \cos{\iorb}\cos\Phi - \cos\eta \sin{\iorb}}.
\end{equation} 
This expression is equivalent to the formula for the polarization angle in the rotating vector model for pulsars \citep{RC69,Meszaros1988,Poutanen2020}.

Once the Stokes flux observed from a surface element is obtained, we can then integrate over the whole SL. 
To account for light bending, we introduce the visibility condition for the surface element $\cos\alpha>0$, as in \citet{Suleimanov20} (see Appendix~\ref{app:a}). 
The misalignment of the NS rotational axis with the accretion disk axis requires one more visibility condition. 
The observer cannot see the parts of the NS covered by the accretion disk, which is mathematically expressed as inequality $\unit{n}\cdot\unit{z}>0$. 
Since $\unit{n}$ is defined in the coordinate system associated with the NS, we need $\unit{z}$ in the same coordinate system:
\begin{equation}
\unit{z} = (\xnew\cdot\unit{z}, \ynew\cdot\unit{z}, \znew\cdot\unit{z}) 
= 
\left(\begin{array}{c} (\cos\iorb-\cos\eta\cos\istar )/\sin\istar \\
- \sin\iorb \sin\eta\sin\Phi/\sin\istar \\ 
\cos\eta \end{array}\right) _*
.
\end{equation}
The second visibility condition $\unit{n}\cdot\unit{z}>0$ then takes the form 
\begin{equation} 
\cos\eta\cos\thetanew +
\sin\eta \sin\thetanew \cos(\Phi_*-\phi) > 0 , 
\end{equation} 
where $\Phi_*$ is the azimuth of the orbital angular momentum vector in the NS frame and 
\begin{eqnarray} 
\left.\begin{array}{l} 
\cos\Phi_* = (\sin\eta\cos\iorb-\cos\eta\sin\iorb\cos\Phi)/\sin\istar, 
 \\
\sin\Phi_* = - \sin\iorb\sin \Phi/\sin\istar . 
\end{array}\right.
\end{eqnarray} 
Thus, for $\thetanew<\pi/2-\eta$, an element is visible for any $\phi$, whereas for $\thetanew>\pi/2+\eta$ it is invisible. In the intermediate regime, $\pi/2-\eta<\thetanew<\pi/2+\eta$, the visibility condition reads $\cos(\Phi_*-\phi)>-\cot\thetanew \cot\eta$, which is equivalent to $|\Phi_*-\phi|<\phi_0\equiv\pi-\arccos(\cot\thetanew \cot\eta)$.

\begin{figure*}
\begin{center}
\includegraphics[width=0.40\linewidth]{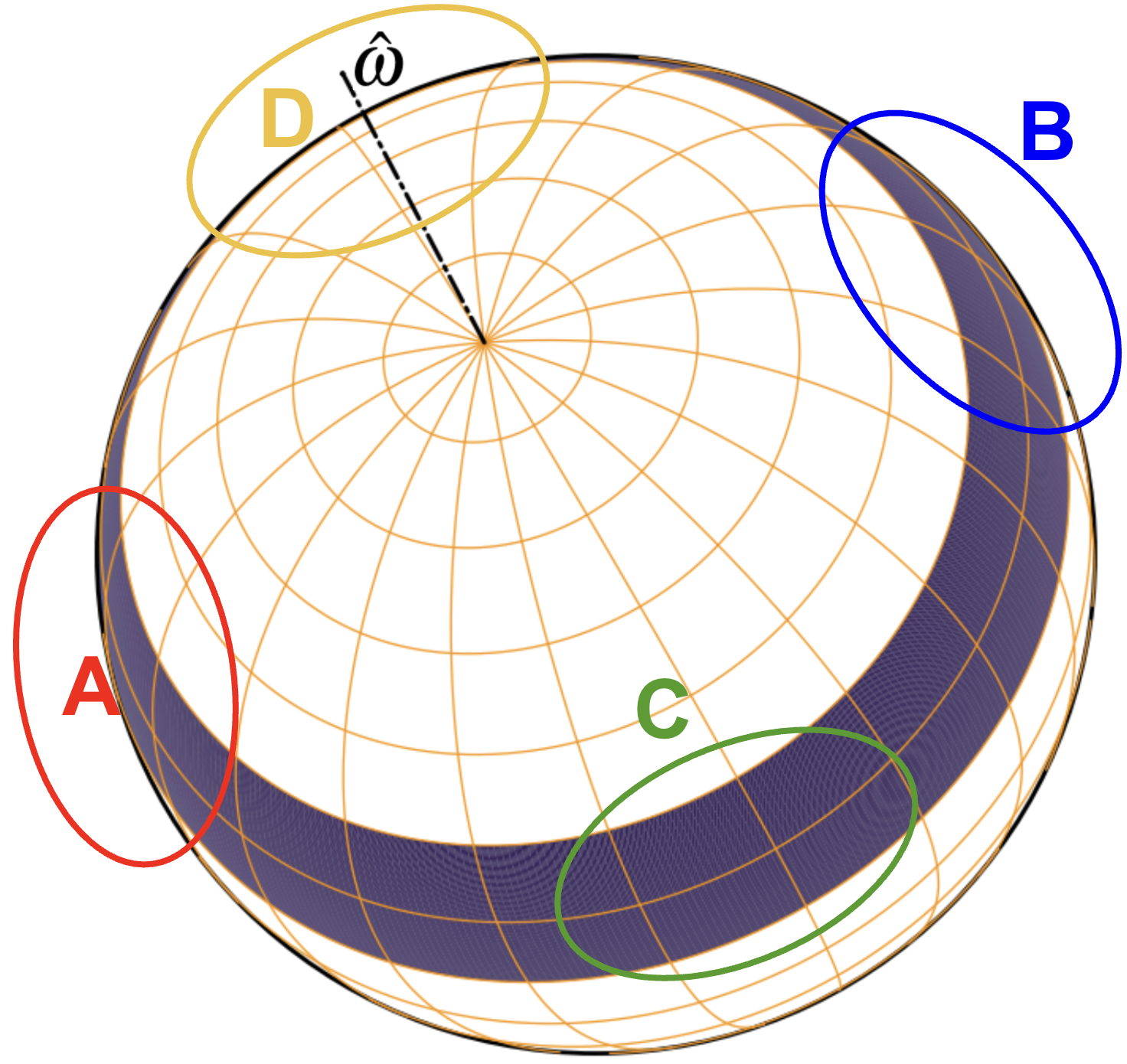} 
\hspace{1cm}
\includegraphics[width=0.40\linewidth]{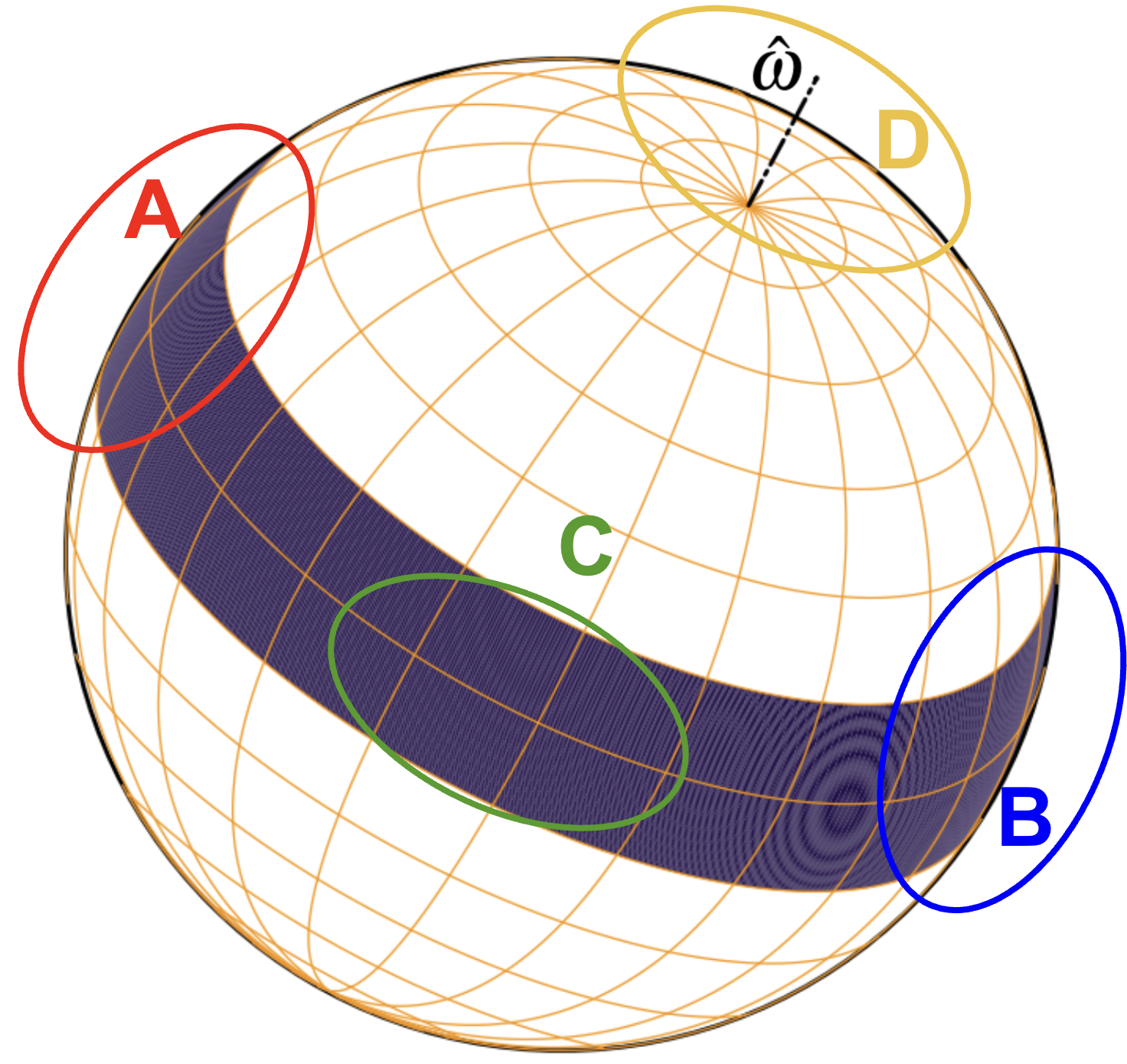}
\caption{A sketch of a SL on the surface of a NS with compactness $u=0.34$ as seen by a distant observer. The matter in the SL flows counterclockwise. The main regions of the SL are shown in different colors: red (zone A) is the flank of the SL with approaching flow; blue (zone B) is the flank of the SL with receding flow; green (zone C) is the main body of the SL; and yellow (zone D) is the back of the SL that is visible in some geometrical configurations due to light bending around the NS. For clarity, the accretion disk is omitted from the sketch.}
\label{fig:scheme}
\end{center}
\end{figure*}

The total polarized flux from the SL is then calculated as follows:
\begin{eqnarray} 
\label{eq:FStokes}
 \bm{F}(E) &=& \left(\begin{array}{c} F_I(E) \\ F_Q(E) \\ F_U (E) \end{array}\right) =  \frac{R^2}{D^2} 
\iint\displaylimits_S
\cos \alpha\ {\cal D} \ \left(\delta\sqrt{1-u}\right)^3 \\ 
&\times& I'_{E'} (\alpha',\thetanew) 
\left(\begin{array}{c} 1 \\ P(\alpha')\cos2\chi_{\rm tot} \\ P(\alpha') \sin2\chi_{\rm tot} \end{array}
\right)\ 
\diff\cos\thetanew\ \diff\phinew ,  \nonumber 
\end{eqnarray} 
where $S$ is the visible NS surface defined by the conditions $\cos \alpha>0$ and $\unit{n}\cdot\unit{z} >0$.
Here $u=R_{\rm S}/R$ is the NS compactness, $R_{\rm S}=2GM/c^2$ is the Schwarzschild radius, $\cal{D}$ is the lensing factor, $I'_{E'}$ is the intensity of escaping radiation in the comoving frame, and $D$ is the distance to the source. 
Finally, we obtain the polarimetric properties of the SL emission: 
\begin{equation}\label{eq:Pobs}
P_{\mathrm{obs}} = \frac{\sqrt{F_{Q}^2+F_{U}^2}}{F_{I}} 
\end{equation} 
and
\begin{equation}\label{eq:chiobs}
\cos(2\chi_{\mathrm{obs}}) = F_{Q}/(P_{\mathrm{obs}}F_{I}) , \quad 
\sin(2\chi_{\mathrm{obs}}) = F_{U}/(P_{\mathrm{obs}}F_{I}) . 
\end{equation}

\section{Results}
\label{sec:Results}

We applied the formalism developed in Sect.~\ref{sec:theory} to study the effects of the misalignment in the system on the observed spectral and polarimetric properties. 
We use dimensionless variables, such as the photon energy:
\begin{equation} 
x = \frac{E}{f_{\rm c}k_{\rm B}T_{\rm eff}},
\end{equation} 
where $k_{\rm B}$ is the Boltzmann constant.
We scaled the observed polarized flux given by Eq.~\eqref{eq:FStokes} to the typical flux at the NS surface $\sigma_{\rm SB}T_{\rm eff}^4$ and the ratio of the distance squared to the emitting area, to introduce the dimensionless luminosity  (Stokes vector): 
\begin{eqnarray}
\label{eq:lum}
 x\, \bm{l}_x\!\! &=&\!\! \frac{D^2}{\Delta R^2}  \frac{E\bm{F}_E}{\sigma_{\rm SB}T_{\rm eff}^4} = \frac{15}{\pi^5} \!\!\iint\displaylimits_S\!\! \cos \alpha\ {\cal D} \ \left(\delta\sqrt{1-u}\right)^4 a_{\rm es}(\alpha') \nonumber  \\
  &\times&  \frac{x'^4}{{\rm e}^{x'}-1}
\left(\begin{array}{c} 1 \\ P(\alpha')\cos2\chi_{\rm tot} \\ P(\alpha') \sin2\chi_{\rm tot}  \end{array}
\right)
\diff\cos\thetanew\ \diff\phinew, 
\end{eqnarray}
where $\Delta$ is the fraction of the NS surface covered by the SL and $x'=x/(\delta \sqrt{1-u})$ is the dimensionless photon energy in the comoving frame (see Appendix~\ref{app:a} for details). 

The results are sensitive to several parameters: the azimuthal angle $\Phi$, the obliquity angle $\eta$, the orbital inclination $\iorb$, the co-latitude of the SL edge $\theta_{\rm min}$, the dimensionless equatorial velocity $\beta_{\rm eq}$ (see Appendix~\ref{app:a}), and the photon energy $x$. 
The parameter combinations used in the simulations presented in this section are listed in Table~\ref{tab:params}.

\begin{figure*}
\begin{center}
\includegraphics[width=1.0\textwidth]{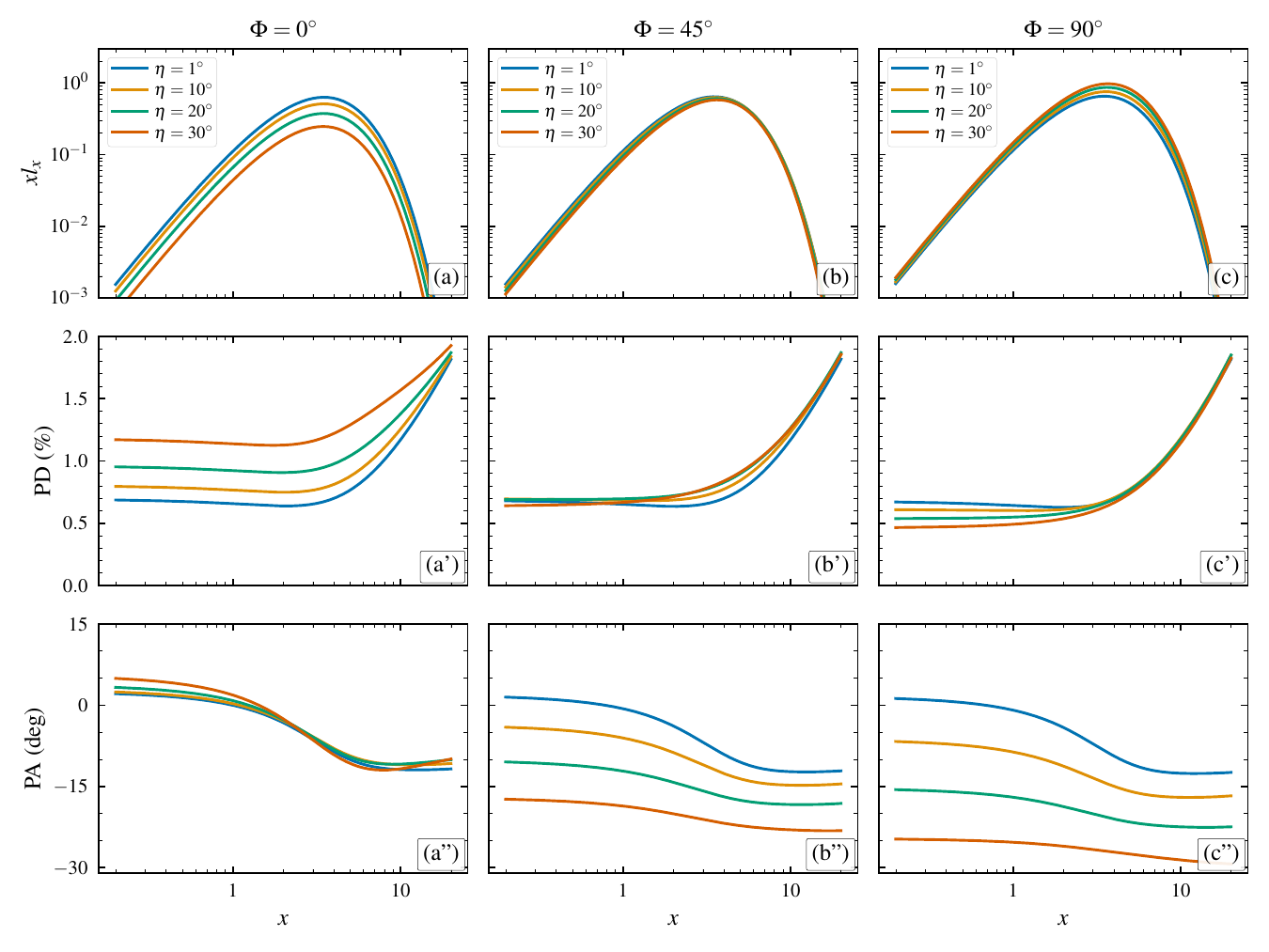}
\caption{Dimensionless luminosity $xl_{x}$, PD, and PA as functions of the dimensionless photon energy, $x$. Fiducial parameter values are listed in Table~\ref{tab:params}. Blue, orange, green, and red lines correspond to obliquity angles $\eta$ of 1\degr, 10\degr, 20\degr, and 30\degr, respectively. Azimuthal angle $\Phi$ is 0\degr\ in panels (a)--(a"), 45\degr\ in panels (b)--(b"), and 90\degr\ in panels (c)--(c"). }
\label{fig:phi_scan}
\end{center}
\end{figure*}

\begin{table}
\caption{List of parameters used in the simulations presented in this section.}\label{tab:params}
    \centering
    \begin{tabular}{ccccc}
    \hline \hline
        Figure&$\Phi$ & $\eta$ & $\iorb$ & $\theta_{\rm min}$ \\ 
         & (deg) & (deg) & (deg) & (deg)  \\ 
        \hline
        Fig.~\ref{fig:phi_scan} (a) & 0 & v & 75 & 60  \\
        Fig.~\ref{fig:phi_scan} (b) & 45 & v & 75 & 60  \\
        Fig.~\ref{fig:phi_scan} (c) & 90 & v & 75 & 60 \\ \hline
        Fig.~\ref{fig:parameter_scan} (a) & 45 & 15 & v & 75  \\ 
        Fig.~\ref{fig:parameter_scan} (b) & 45 & 15 & 75 & v \\
        Fig.~\ref{fig:parameter_scan} (c) & v & 15 & 75 & 75  \\ \hline
        Fig.~\ref{fig:contours} (a) & $-$ & $-$ & 30 & $90 - \eta$  \\ 
        Fig.~\ref{fig:contours} (b) & $-$ & $-$ & 50 & $90 - \eta$  \\ 
        Fig.~\ref{fig:contours} (c) & $-$ & $-$ & 70 & $90 - \eta$  \\ \hline
        Fig.~\ref{fig:contours_cos} (a) & $-$ & 15 & $-$ & 75  \\ 
        Fig.~\ref{fig:contours_cos} (b) & $-$ & 30 & $-$ & 60  \\ 
        \hline
    \end{tabular}
    \tablefoot{Parameters varied along the plot axes are indicated by a "$-$" sign. The letter "v" indicates that multiple values of the parameter are used and specified in the corresponding plot legend. The equatorial velocity is fixed at $\beta_{\rm eq}=0.35$ for all studies presented in this section. The energy is fixed at $E = 5\rm~keV$ for Figs.~\ref{fig:contours}~and~\ref{fig:contours_cos}. }
\end{table}

In Fig.~\ref{fig:scheme}, a sketch of a SL on the surface of an misaligned WMNS with compactness $u=0.34$ is shown. The main regions of the SL are marked in different colors. These regions make significantly different contributions to both the luminosity and the polarimetric properties of the SL emission. Regarding the luminosity, Zone C accounts for the majority of the emission, Zone A contributes at higher energies, Zone B contributes at lower energies, and Zone D is only visible in some cases. For a plane-parallel slab, the polarimetric properties depend on the angle between the slab surface normal and the direction to the observer: the larger this angle, the larger the PD. This means that Zones A (approaching flow on the flank of the SL) and B (receding flow on the other flank of the SL) emit more strongly polarized light in the direction of the observer than Zone C (main body of the SL), whereas emission from Zone C is mostly unpolarized. Zone D (back of the SL) has a PD similar to those of Zones A and B, although the PA is different: since the emission from a plane-parallel slab is polarized in the plane of the slab surface, the PA of the emission from Zone D is almost orthogonal to those of Zones A and B.  

\subsection{Tilt effects on the polarization signatures} 

\begin{figure*}
\begin{center}
\includegraphics[width=1.\textwidth]{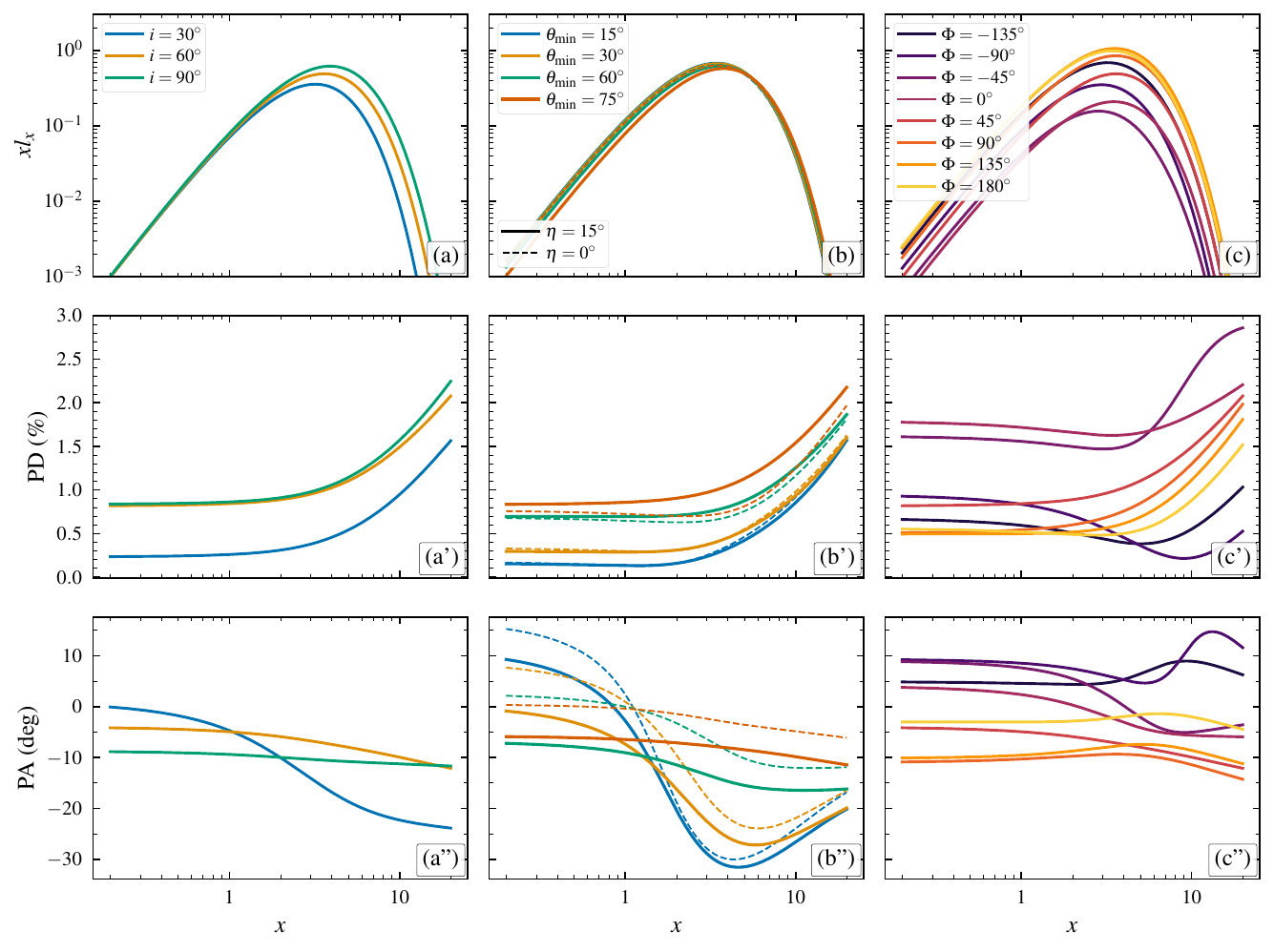}
\caption{Dimensionless luminosity $xl_{x}$, PD, and PA as functions of the dimensionless photon energy, $x$. Fiducial parameter values are listed in Table~\ref{tab:params}. Panels (a)--(a''): the orbital inclination is varied; blue, orange, and green lines correspond to 30\degr, 60\degr, and 90\degr, respectively. Panels (b)--(b''): the co-latitude of the SL edge $\theta_{\rm min}$ is varied; blue, orange, green, and red lines correspond to 15 \degr, 30\degr, 60\degr, and 75\degr, respectively. Solid lines correspond to $\eta=15\degr$, while dashed lines correspond to the aligned case with $\eta=0\degr$ from \citetalias{Bobrikova25}. Panels (c)--(c''): the azimuthal angle $\Phi$ is varied; the lines, colored according to the \texttt{inferno} palette, correspond to $-135\degr$, $-90\degr$, $-45\degr$, 0\degr, 45\degr, 90\degr, 135\degr, and 180\degr, respectively.}
\label{fig:parameter_scan}
\end{center}
\end{figure*}

In this section, we study the effect of the azimuthal angle $\Phi$ and the obliquity $\eta$ on the energy dependence of luminosity, PD, and PA.
We chose a rather wide SL with $\theta_{\rm min}=60\degr$ and $\theta_{\rm max}=120\degr$, so that varying $\eta$ up to 30\degr\ would not break the continuity of the accretion flow. We fixed the inclination at $\iorb=75\degr$ and the SL equatorial velocity to $\beta_{\rm eq}=0.35$; for these parameters, a relatively high PD is expected \citepalias{Bobrikova25}. The combination pf parameters used for each case is listed in Table~\ref{tab:params}.

In Fig.~\ref{fig:phi_scan}, we consider three different values of the azimuthal angle, $\Phi$, of $0\degr,~45\degr,~\rm and~90\degr$. For each case, we show the dimensionless luminosity, PD, and PA for four different values of the misalignment angle $\eta$.
In panels (a)--(a''), the observer is in the plane formed by the NS and disk axes (we denote this plane by $\tau$). The dimensionless luminosity decreases with the obliquity angle $\eta$, while PD increases with $\eta$. In all cases, the emission is polarized perpendicular to the disk. 

The increase in PD is caused by the partial coverage of Zone C by the accretion disk. As $\eta$ increases, the contribution of Zone C to the total flux decreases, while the contributions of Zones A and B become more important. The reduced contribution of Zone C therefore leads to an increase in the net polarization. At higher energies, the contribution of Zone A becomes dominant. This explains both the increase in PD at higher energies and the slight change in PA with energy.

In panels (c)--(c'') of Fig.~\ref{fig:phi_scan}, the plane $\tau$ is orthogonal to the $xz$-plane, in which the observer vector $\unit{k}$ lies. In this geometry, changes in obliquity have a weaker effect on PD. The main effect of changing the obliquity appears at lower energies, where PD decreases with increasing obliquity. 
The PA, on the other hand, changes significantly with $\eta$, with $\mathrm{PA}\approx-\eta$. 

In this geometry, the accretion disk partially covers Zone B. The contribution of Zone B is most prominent at lower energies, where PD decreases with increasing obliquity. As $\eta$ increases, a larger fraction of Zone B is covered, reducing the contribution of Zone B to the total emission. Since the emission from Zone B is more strongly polarized than that from Zone C, this leads to a decrease in PD. The emission then comes mainly from Zones A and C, and PA is determined by the orientation of Zone A with respect to the accretion disk plane. Since this orientation depends on $\eta$, this produces a correlation between the predicted PA and obliquity $\eta$. 
 
The case of $\Phi=45\degr$ in panels (b)--(b'') is less straightforward. It shows intermediate PD and a complex PA dependence on $\eta$, as expected for a combination of the behaviors seen in panels (a)--(a'') and (c)--(c’’). In this geometry, parts of both Zones B and C are covered by the accretion disk, so the mechanisms present in the two previous cases are present simultaneously. For PD, these two effects together result in little dependence of PD on $\eta$. For PA, the orientation of Zone A with respect to the accretion disk plane remains a dominant factor, as in panel (c''), although the PA-$\eta$ relation is mildly affected by the different azimuthal angle $\Phi$.

\begin{figure*}
\begin{center}
\includegraphics[width=0.85\textwidth]{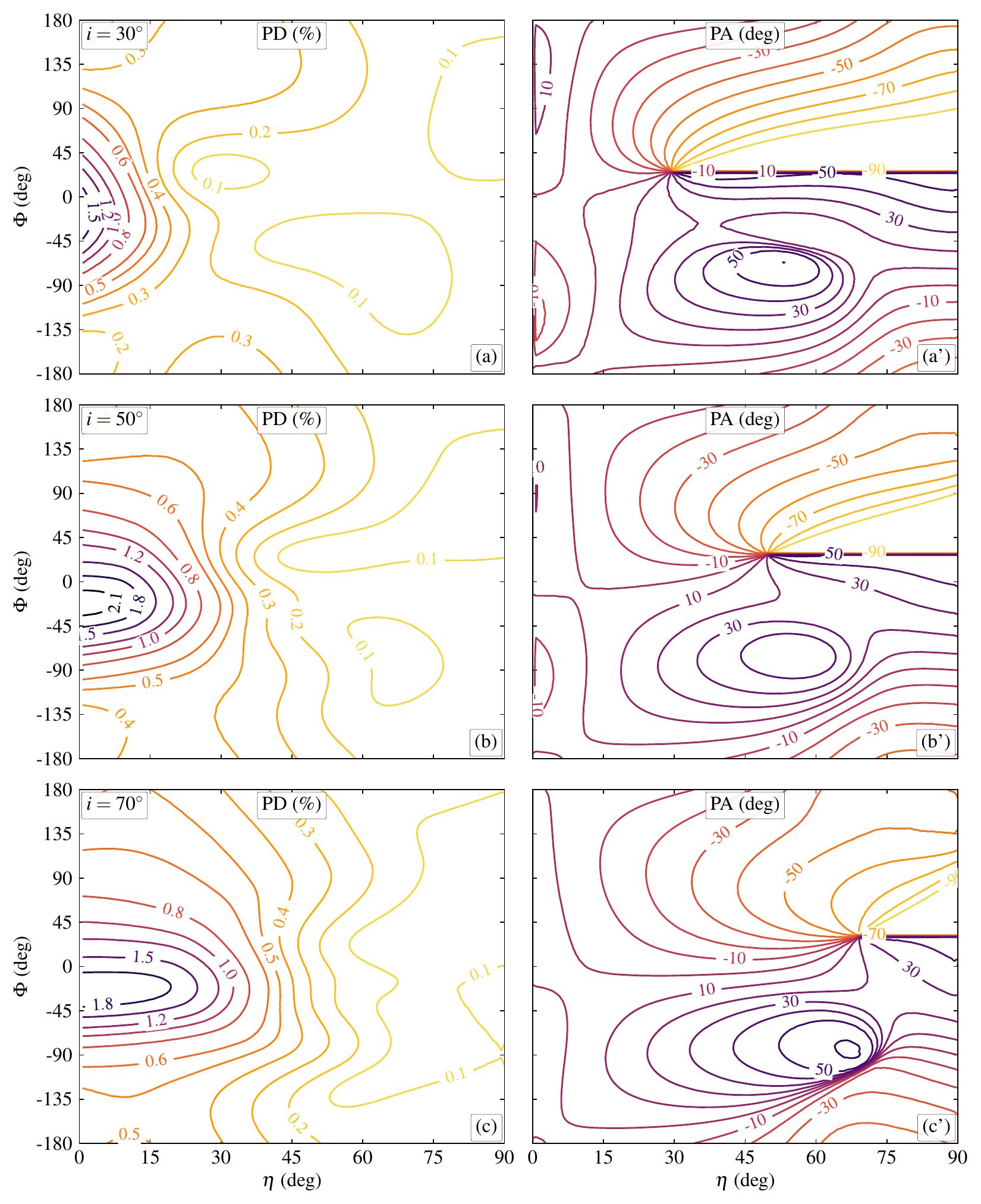}
\caption{Contour plots of PD and PA as functions of obliquity $\eta$ and azimuthal angle $\Phi$. Panels (a)--(a'): orbital inclination $\iorb=30\degr$; panels (b)--(b'): $\iorb=50\degr$; panels (c)--(c'): $\iorb=70\degr$. Remaining parameter values are listed in Table~\ref{tab:params}.}
\label{fig:contours}
\end{center}
\end{figure*}

\subsection {Parameter space study}

Next, we considered the effects of changing the orbital inclination $\iorb$, the co-latitude of the SL edge $\theta_{\rm min}$, and the azimuthal angle $\Phi$ for a fixed obliquity $\eta=15\degr$. The full set of parameters for each panel is given in Table~\ref{tab:params}. The results are shown in Fig.~\ref{fig:parameter_scan}. 

Panels (a)--(a'') illustrate the effect of changing the orbital inclination $i_{\rm{orb}}$. 
The results show the standard trend of PD increasing with orbital inclination and PA varying slightly around the direction orthogonal to the disk plane \citepalias{Bobrikova25}. In the case of $\iorb=30\degr$, all four Zones contribute to the flux, while for the other two cases, Zone D does not. Panel (a) shows a shift in the peak energy, which is associated with the Doppler effect. For a fixed velocity $\beta_{\rm eq}$, the Doppler effect is stronger at higher inclinations, because the projection of the velocity onto the direction to the observer is larger at higher inclinations in the current configuration of the system.

In panels (b)--(b''), we showed the impact of the SL width by varying the edge co-latitude $\theta_{\rm min}$. For comparison, we also showed the corresponding results for an aligned star with dashed lines \citepalias{Bobrikova25}. Since dimensionless luminosity is normalized to the surface area of the SL, panel (b) shows almost no effect of $\theta_{\rm min}$ on $xl_x$. 
Panel (b') shows that, similarly to the aligned case \citepalias{Bobrikova25}, the narrower the SL, the higher the PD. In the case of $\theta_{\rm min}=15\degr$, most of the NS surface is covered by the SL, and the geometry of the SL approaches a sphere, which contributes to depolarization \citepalias{Bobrikova25}. This effect weakens with increasing $\theta_{\rm min}$. 

The misalignment in this case contributes to a minor increase in PD for $\theta_{\rm min}=75\degr$ and $\theta_{\rm min}=60\degr$, while the emission from a wider SL seems almost unaffected by the misalignment. Similarly, PA is mildly affected by the misalignment, with the difference being more noticeable for narrower SLs with $\theta_{\rm min}=75\degr$ and $\theta_{\rm min}=60\degr$.
The reason behind this change with respect to the case of $\eta=0\degr$ again lies in the partial covering of Zone C by the accretion disk and the slight asymmetry introduced by observing the system from $\Phi=45\degr$. This effect is less present for a wider SL, where emission is depolarized.

\begin{figure*}
\begin{center}
\includegraphics[width=0.85\textwidth]{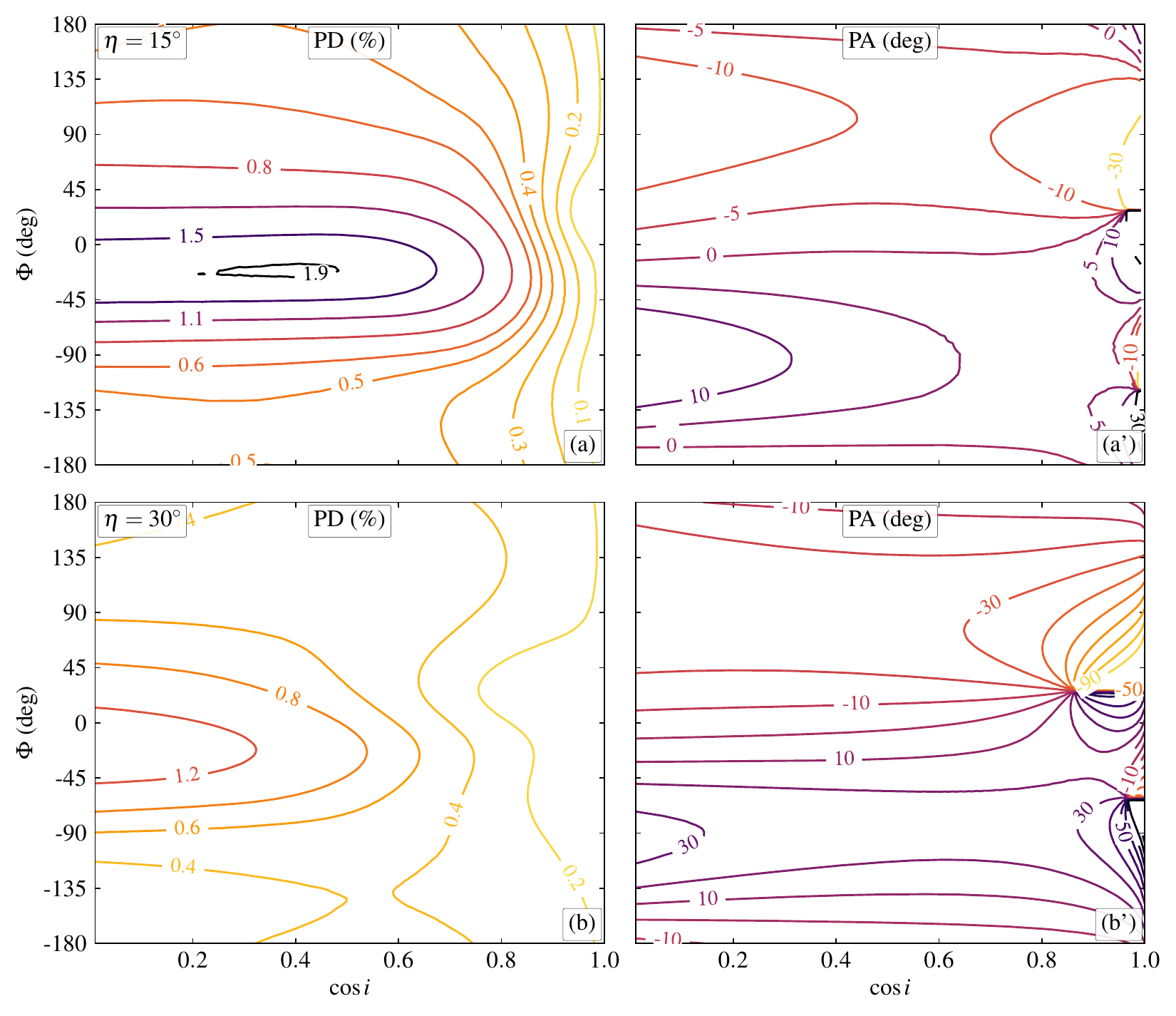}
\caption{Contour plots of PD and PA as functions of the azimuthal angle $\Phi$ and the cosine of the orbital inclination $\cos i$. Panels (a)--(a'): obliquity $\eta=15\degr$; panels (b)--(b'): $\eta=30\degr$. 
Remaining parameter values are listed in Table~\ref{tab:params}.}
\label{fig:contours_cos}
\end{center}
\end{figure*}

In panels (c)--(c''), we show the effect of varying the azimuthal angle $\Phi$. The dimensionless luminosity shown in panel (c) strongly depends on $\Phi$. For $\Phi=0\degr$ or $180\degr$, the problem is axially symmetric, similar to the case presented in Fig.~\ref{fig:phi_scan}, panels (a)--(a''). Zones A and B contribute equally to the flux. However, for $0\degr<\Phi<180\degr$, Zone B is partially or completely covered by the accretion disk, and hence the peak energy increases. For $-180\degr<\Phi<0\degr$, the situation is reversed: Zone A is partially or completely covered by the accretion disk, and the peak energy decreases. Since the luminosity is normalized to the full surface area of the SL, not to the visible part, the peak luminosity also changes significantly, increasing by an order of magnitude between the case of $\Phi=0\degr$, where Zone C is covered by the accretion disk, and the case of $\Phi=180\degr$, where a larger fraction of Zone C is visible, since the star is tilted in the direction opposite to the observer. 

We note that for $\Phi=0\degr$ and $\Phi=-45\degr$, the PD is higher than in other cases. The PA varies within a range of approximately 20\degr\ as $\Phi$ changes, first slowly decreasing as $\Phi$ increases from $-45\degr$ to $135\degr$, and then slowly increasing for the other hemisphere. This is caused by the fact that PA mostly depends on the minimal  co-latitude that is covered by the SL, which changes slowly with the azimuthal angle in the misaligned case. For a larger obliquity angle $\eta$, a larger variation in PA is expected.

\subsection{Monochromatic polarization as a function of orientation} \label{sec:contour}

After studying the energy dependence of spectropolarimetric properties, we fixed the photon energy and calculated PD and PA for all possible combinations of obliquity $\eta$ and azimuthal angle $\Phi$. We linked the co-latitude of the SL edge to the obliquity angle through the relation $\theta_{\rm min}=90\degr-\eta$, so that there is never a gap between the SL and the accretion disk. In this configuration, the SL is wider for systems with larger misalignment. We kept the same equatorial flow velocity $\beta_{\rm eq}=0.35$. To characterize the average PD and PA in the \ixpe energy range, we chose to study monochromatic polarization at $E=5~\rm keV$. In the previous section, we used the dimensionless energy $x$, and since the energy is now expressed in keV, we also had to fix the effective temperature $T_{\rm eff}$ and the color correction $f_{\rm c}$ for the SL. Based on spectroscopic studies of WMNSs, we adopted $f_{\rm c}~T_{\rm eff}=2$~keV. As before, all the parameter values for each simulation are listed in Table~\ref{tab:params}.

In Fig.~\ref{fig:contours}, the results are shown for three different orbital inclinations, $\iorb$, of 30\degr, 50\degr, and 70\degr. The case of the lowest orbital inclination presented in panel (a) shows $\rm PD<0.5\%$, except for the region of $\eta<15\degr$. For small $\eta$, the width of the SL plays a crucial role, significantly increasing the polarization. Panels (b) and (c) show that, for moderate or high inclination, PD can remain higher than 1\% even at larger values of $\eta$, which in this simulation correspond to a rather wide SL. We note that PD reaches its maximum for $-45\degr<\Phi<0\degr$. Contrary to the results shown in panels (a'), (b'), and (c') of Fig.~\ref{fig:phi_scan}, PD is higher for smaller obliquity angles in this simulation. Here, $\eta$ is connected to the SL width, and the effect of the SL width on PD dominates the behavior, while in Fig.~\ref{fig:phi_scan}, the SL width was fixed. The PA is strongly influenced by the obliquity, as seen in panels (b'') and (c'') of Fig.~\ref{fig:phi_scan}. The behavior of the polarimetric properties shows a common topology for different inclinations; for instance, the PD maxima occur in similar regions in panels (b) and (c). PA depends strongly on both $\Phi$ and $\eta$ at all considered inclinations. 

Next, we investigated the probability of detecting the PD higher than the threshold for a fixed obliquity $\eta$. 
For this study, we assumed a flat distribution for both $\cos \iorb$ and $\Phi$. 
The results are shown in Fig.~\ref{fig:contours_cos}. 
We show the constant PD contours for two values of obliquity, $\eta$, of 15\degr\ and 30\degr. Contrary to panels (a') and (b') of Fig.~\ref{fig:phi_scan}, where the highest PD was expected for $\eta=30\degr$, here we observe the decrease of PD with misalignment; this is due to the fact that SL is wider for $\eta=30\degr$, since we set $\theta_{\rm min}=90\degr-\eta$. The calculated probabilities of observing a PD higher than the threshold value are given in Table~\ref{tab:prob}. We considered six obliquity values and four PD levels. We did not consider $\eta>30\degr$, since PD drops rapidly at higher obliquity due to an increase in the SL width. 
Since the probability distribution of $\eta$ is likely not flat, to understand the true probability of observing the system at a given level of PD, we need to convolve the reported probabilities with the probability distribution of $\eta$ in the WMNS population. We can, however, conclude that only for a few values of $\eta$ does the probability of observing a PD higher than 1\% exceed 20\%, and the 2\% PD level is rarely reached, with a probability exceeding 1\% only for $\eta=5\degr$.

\begin{table} 
\caption{Probability (in \%) of detecting PD higher than the threshold 
for different obliquity values.
}\label{tab:prob}
\centering
\small
\begin{tabular}{ccccccc}
\hline\hline
\multirow{2}{*}{{PD} (\%)} & \multicolumn{6}{c}{$\eta$ (deg)} \\ 
  \cline{2-7}
& $5$ 
& $10$
& $15$
& $20$
& $25$
& $30$
\\
\hline
$\ge0.5$ & 39  & 50  & 48  & 55  & 46  & 36  \\
$\ge1.0$ & 28  & 25  & 21  & 17  & 12  & 7.8   \\
$\ge1.5$ & 15  & 12  & 9.2  & 6.1  & 3.0  & 0.4    \\
$\ge2.0$ & 2.9  & 0.8  & 0.0  & 0.0  & 0.0  & 0.0  \\

\hline
\end{tabular}
\end{table}

\section{Discussion} \label{sec:IXPE}

The results presented in Sect.~\ref{sec:Results} show several important trends. Regarding the spectrum, the main result is that misalignment affects the spectral shape, an effect that should be taken into account when performing the spectral decomposition. For the PD, we note that breaking the axial symmetry increases the expected PD with respect to the aligned case presented in \citetalias{Bobrikova25}, with PD exceeding 2\% for some parameter combinations (see Table~\ref{tab:prob} for examples). Regarding the PA, we conclude that misalignment in the system allows the polarization plane to take any orientation. 

Before comparing the model predictions with the observational results, it is important to discuss the possibility of misalignment in a WMNS system. When an X-ray binary is formed following the collapse of a normal star into a compact object, the rotation axis of the compact object is not required to be aligned with the orbital axis \citep{2006ApJ...639.1007W}. The accretion process reduces the obliquity, since angular momentum is transferred from the disk to the NS via the accretion flow. The timescale for the alignment process depends on the mass accretion rate, magnetic field strength, rotation rate of the NS, and other parameters. 

Theoretical expectations for the alignment time in binaries are discussed in the literature for several subclasses of sources: for instance, \citet{1987A&A...173..279D} estimated the alignment time for millisecond pulsars at $\approx10^6$~years, while \citet{Abolmasov2024} suggests a significantly shorter alignment time for strongly magnetized NSs. \citet{Martin2008} estimated an alignment time of $10^7$~years for the black hole X-ray binary GRO~J1655$-$40 is reported. However, no theoretical expectations have been derived for WMNSs. Although most of these objects are considered to be rather old \citep[e.g. the age estimate of Cyg~X-2 is $4\times10^8$~yr;][]{1999MNRAS.309..253K}, some are known to be considerably younger, and possibly young enough to remain misaligned. For instance, the age of \mbox{Sco~X-1} is estimated to be $7\times10^6$~yr \citep{2025ApJ...980...51K}, while \mbox{Cir~X-1} is less than $5\times10^3$~yr old \citep{2013ApJ...779..171H}. 
The alignment time can also be extended by additional factors related to binary evolution and by events occurring during it.
For instance, if the accretion process has been episodic, or if the binary system is particularly wide, the accretion-driven alignment could take a significantly longer time \citep{2025ApJ...978...80L}. 

\mbox{Cir~X-1} is the first and most compelling candidate for a potentially misaligned WMNS. Apart from being the youngest known WMNS system, \mbox{Cir~X-1} is known to have precessing jets \citep{2025MNRAS.544L..37C} and a high eccentricity of 0.4--0.45 \citep{2007MNRAS.374..999J, 2016MNRAS.456..347J}, with even higher values reported in earlier works, e.g., 0.94 in \citet{1999MNRAS.310.1165T}. The presence of a kick during the supernova event in this system has been discussed as a possible explanation for both the jets and the eccentricity, and with an age of less than $10^4$~yr, there may not have been sufficient time for the NS rotational axis to align with the orbital axis through accretion. 

The \ixpe observation of \mbox{Cir X-1} in August 2023 revealed strong PA variability with spectral hardness and source state, as well as a PA that was misaligned with both the assumed disk plane and the jet axis. Obliquity in the system has been suggested as one of the possible explanations for the observed polarimetric behavior. \citet{Rankin2024} suggest that, based on the temporal evolution of PA, \mbox{Cir~X-1} exhibited different accretion-flow geometries during the observations, with the geometry during the part of observation labeled P3 including an SL. During that part of the observation, \mbox{Cir~X-1} had a PD of 1.4\% and a PA$=-12\degr$ neither orthogonal nor aligned with the jet position angle reported to be between $-40\degr$ and $-70\degr$. If we assume the jet direction to be orthogonal to the accretion disk plane, the observed PA would correspond to $30\degr$ and $60\degr$ in our simulations. 
As shown in Fig.~\ref{fig:contours}, such a PA is expected for an azimuthal angle $\Phi$ between $\approx-135\degr$ and $\approx-45\degr$ and an obliquity $\eta$ larger than $\approx30\degr$. More precise constraints on $\eta$ are difficult to derive due to the uncertainty in the PA. We note that the exact numbers also depend on the inclination of the system and other parameters that are assumed rather than inferred, in the simulations. 
Accounting for both PD and PA does not tighten the constraints; Fig.~\ref{fig:contours} illustrates that PD does not reach 1.4\% within the stated ranges of obliquity and azimuthal angles in our simulation, at least for the selected set of parameters. We can, however, conclude from Fig.~\ref{fig:contours} that higher inclinations are favored, since PD in the selected region rises with inclination. This could be interpreted as a hint of a relatively high inclination of the system. 
However, we note that the PD prediction is strongly affected by assumptions regarding the SL structure and composition; while we consider only scattering off electrons, introducing ions and accounting for absorption in the SL could significantly increase the expected PD \citep[][]{2024ApJ...964...77R}. 
Given these uncertainties, we do not derive any tighter constraints on the \mbox{Cir~X-1} geometry from the PD predictions.

Apart from \mbox{Cir~X-1}, a few other systems have exhibited signs of misalignment based on \ixpe observations. 
For instance, in \mbox{GX~5$-$1}, a swing by $\approx 50\degr$ between the PA of the disk and the Comptonized components was reported \citep{Fabiani24}. If we assume the disk to be polarized in the plane of the disk \citep{LVP22}, corresponding to a $90\degr$ PA in this paper, then the Comptonized component is polarized in the direction of $\approx 40\degr$, which can be reproduced by our model.  
Although both of these systems are considered old, with an approximate age of $10^8$~yr, little is known about the duration of their active accretion phase. 
We neither favor nor rule out misalignment as a possibility in these systems; rather, we provide a model to investigate the polarimetric properties of these objects under the assumption of misalignment. 
We look forward to studying the results of further spectropolarimetric observations of these systems. 

\section{Summary}
\label{sec:Summary}


We developed a model for the spectropolarimetric properties of the emission of a SL on a tilted WMNS, whose rotational axis is not aligned with that of the binary system. 
We investigated how luminosity, PD, and PA depend on energy and on the geometrical parameters of the system, such as the obliquity angle $\eta$, the azimuthal angle $\Phi$, the inclination of the binary system $\iorb$, and the co-latitude of the SL edge $\theta_{\rm min}$.

We conclude that the expected polarimetric properties of the SL emission depend strongly on the geometrical parameters of the system. First, we find that breaking the axial symmetry contributes to an increase in the PD of the SL emission. Additionally, we show that misalignment in the system could explain why the observed PA is neither aligned with nor orthogonal to the assumed disk plane or binary rotational axis. Consistent with the aligned case studied in \citetalias{Bobrikova25}, PD increases with energy and inclination and decreases with increasing SL width. We find that a higher PD is predicted for azimuthal angles in the range $\Phi=-45\degr$--$0\degr$. The effect of the obliquity angle strongly depends on the SL width, which is connected to the mass accretion rate \citep{IS99}. For luminous systems with a wider SL, PD is higher for larger values of $\eta$, as seen in Fig.~\ref{fig:phi_scan}; in less luminous systems, where the width of the layer is connected to the obliquity as shown in Sect.~\ref{sec:contour}, PD is higher for lower obliquity.

In recent years, observations of several systems have shown indications of misalignment, and in this paper we predict their polarimetric properties under the assumption of misalignment. The formalism derived here can be used to interpret existing and upcoming spectropolarimetric observations of WMNSs. 
The misalignment scenario can be realized in cases in which transfer of angular momentum from the accretion disk has not significantly altered the orientation of the NS rotational axis. The most natural candidates for this scenario are young systems such as \mbox{Cir~X-1}. Older systems like \mbox{GX~13+1} and other WMNSs require episodic mass transfer or additional mechanisms to reduce the effect of the accretion process on the NS spin; in such cases, a non-negligible obliquity may persist even in older systems.

\begin{acknowledgements}
This research was supported by the grant 00240328 of the Finnish Cultural Foundation (AB), by the Research Council of Finland Centre of Excellence in Neutron-Star Physics (grant 374064) and grants 355672 and 372881 (AV).
\end{acknowledgements}

\bibliographystyle{yahapj}
\bibliography{ns}

\begin{appendix}

\section{Polarized emission from an aligned spreading layer}\label{app:a}

To calculate the polarized emission from the SL, we first derived an equation for the polarized flux coming from a small surface element $\diff F_{E}$, and then integrated that emission over the whole visible surface of the SL in the coordinate system associated with the NS rotation axis. 
Main equations have been derived in \citet{Poutanen2020} and \citet{Suleimanov20}. 
For the assumed spherical NS, the surface normal $\unit{n}$ coincides with the radius-vector of the element $\unit{r}$ and thus the angle $\psi$ between $\unit{n}$ and the direction to the observer $\unit{k}$ is
\begin{equation} 
\cos\psi \equiv \unit{k}\cdot\unit{n} = \cos \istar \cos\thetanew + \sin \istar \sin\thetanew \cos\phinew ,
\end{equation} 
where $\phinew = 0$ when the element was closest to the observer. 
The emission angle $\alpha(\psi)$ (i.e., the angle to the surface normal in the external frame) was calculated using \citet{B02} approximation 
\begin{equation}\label{eq:alpha}
\cos\alpha(\psi)\equiv\unit{k}_0\cdot\unit{n}\approx 1-(1-u)(1-\cos\psi) , 
 \end{equation} 
with the visibility condition for the surface element being $\cos\alpha>0$. 
Here, $u=R_{\rm S}/R$ is the NS compactness, and $\unit{k}_0$ is the unit vector of the photon momentum close to the surface, 
\begin{equation}\label{eq:k0}
\unit{k}_0 = \frac{\sin \alpha\ \unit{k} + \sin{(\psi - \alpha)}\ \unit{n}}{\sin{\psi}}.
\end{equation} 

Similarly to \citetalias{Bobrikova25}, we defined the velocity of the material in the SL by the direction of the unit vector $\unit{v} = (-\sin\phinew, \cos\phinew, 0)_*$ and the amplitude (in units of speed of light) as a function of latitude 
\begin{equation} \label{eq:velocity}
\beta(\thetanew) = \beta_{\rm eq} \sin\thetanew = \frac{A \ v_{\rm kepl}}{c} \sin\thetanew = A \ \sqrt{\frac{u}{2 \ (1-u)}} \sin\thetanew ,
\end{equation} 
with $A<1$ being a proportionality coefficient that scales equatorial velocity relative to the Keplerian velocity $v_{\rm kepl}$ in the Schwarzschild metric \citep{Luminet1979}. 
The angle between the velocity vector and the photon direction is then \citep{PG03}: 
\begin{equation}\label{eq:cosxi}
\cos\xi = \unit{v} \cdot \unit{k}_0 = \frac{\sin\alpha}{\sin\psi}\ \unit{v} \cdot \unit{k}= - \frac{\sin\alpha}{\sin\psi} \sin \istar \ \sin\phinew\ .
\end{equation} 

We calculated the observed flux at energy $E$ from the surface element with area $\diff S = R^2\diff\cos\thetanew \diff\phinew$ as follows:
\begin{equation} 
\diff F_E = I_E \diff\Omega_{\rm obs} = {\cal D} \cos \alpha \frac{\diff S} {D^2} , 
\end{equation} 
where $\cal D$ is the lensing factor that is equal to one in the Beloborodov approximation. 

To account accurately for the relativistic effects of the matter rotation, we moved to the frame co-rotating with the SL matter. We considered the SL element to be an electron-dominated semi-infinite slab \citep{Cha60, Sob63} so that we could compute the angular distribution and polarization of the escaping radiation. 
We assumed the local intensity $I'_{E'}(\alpha',\thetanew)$ to have a diluted blackbody shape, same as in \cite{Suleimanov20} and \citetalias{Bobrikova25}, given by Eq.~\eqref{eq:dilutbb}. 
The angle $\alpha'$ made by the photon momentum with the surface normal in the co-rotating frame is related to the emission angle $\alpha$ defined in Eq.~\eqref{eq:alpha} as \citep{PG03} 
\begin{equation} 
\cos\alpha' = \delta \cos\alpha,
\end{equation} 
where $\delta$ is the Doppler factor
\begin{equation}\label{eq:Doppler_sph}
\delta = \frac{1}{\gamma(1-\beta\cos\xi)} 
\end{equation} 
and $\gamma$ is the Lorentz factor 
\begin{equation} \label{eq:Lorentzgamma}
\gamma(\thetanew)=\frac{1}{\sqrt{1-\beta^2(\thetanew)}}. 
\end{equation} 

The local intensity $I'_{E'}$ can depend on co-latitude $\thetanew$ but we consider here the effective temperature and the color correction factors to be constant throughout the SL. 
The observed intensity is related to the local one: 
\begin{equation} \label{eq:IE_EE}
I_{E} = \left (\frac{E}{E'}\right )^3 I'_{E'} (\alpha',\thetanew) = \left( \delta \sqrt{1-u}\right)^3 I'_{E'}.
\end{equation} 
The total observed flux coming from the surface element was calculated as 
\begin{equation} 
\diff F_{E} = \frac{R^2}{D^2} \cos \alpha\ {\cal D} \ \left(\delta\sqrt{1-u}\right)^3 I'_{E'} (\alpha',\thetanew) \,\diff\cos\thetanew\, \diff\phinew, 
\end{equation} 
where $D$ is the distance to the source. 

Polarimetric properties of the SL emission can be described by the Stokes vector:
\begin{equation}
\diff \bm{F}(E)= \left(\begin{array}{c} \diff F_I(E) \\ \diff F_Q(E) \\ \diff F_U (E) \end{array}\right) = \diff F_E \left(\begin{array}{c} 1 \\ P\cos2\chi \\ P \sin2\chi \end{array}
\right) ,
\end{equation} 
where $P$ is the PD of the escaping radiation, and $\chi$ is the PA. 
We calculated PD using the approximation from \citet{VP04} for a semi-infinite electron-scattering-dominated atmosphere \citep{Cha60}:
\begin{equation} \label{eq:ChandrasekharPD}
P=P_{\mathrm{es}} (\alpha') = - \frac{1-\cos\alpha'}{1+3.582 \cos\alpha'} \ 0.1171, 
\end{equation} 
and the PA with the formulae derived by \citet{Poutanen2020}:
\begin{equation} \label{eq:tanchi_rel_bend}
\tan \chi = 
\frac{ \sin \thetanew \sin \phinew + \beta \ A } 
{-\sin \istar \cos \thetanew + \cos \istar \sin \thetanew \cos\phinew - \beta\sin\phinew \ C }, 
\end{equation}  
where 
\begin{eqnarray} \label{eq:aux_rel_bend}
A&=& \frac{\sin\psi}{\sin\alpha} B+ \frac{\cos\alpha-\cos\psi}{\sin\alpha \sin\psi} (\cos\phi-B\cos\psi) , \nonumber \\ 
B&=& \sin \istar \sin \thetanew+\cos \istar \cos \thetanew \cos\phinew, \\
C&=& \frac{\sin\psi}{\sin\alpha}\cos \theta + \frac{\cos\alpha-\cos\psi}{\sin\alpha \sin\psi} (\cos \istar -\cos \thetanew \cos\psi) . \nonumber
\end{eqnarray}
These equations account for the effects of rotation of polarization plane because of relativistic motion as well as for light bending in the Schwarzschild metric. 

\end{appendix}
\end{document}